\documentclass[preprintnumbers]{revtex4}
\UseRawInputEncoding
\usepackage{amssymb}
\usepackage{amsmath}
\usepackage{graphicx}
\usepackage{dcolumn}
\usepackage{bm}
\usepackage{subfigure}
\usepackage{color}

\begin{document}

\title{Thermodynamics of Barrow Einstein-power-Yang-Mills AdS black hole in the restricted phase space}
\author{Yun-Zhi Du$^{1,2}$, Hui-Hua Zhao$^{1,2}$\footnote{the corresponding author}, Yang Zhang$^{1,2}$,  and Qiang Gu$^{1,2}$}
\address{$^1$Shanxi Datong University, Datong 037009, China\\
$^2$State Key Laboratory of Quantum Optics and Quantum Optics Devices, Shanxi University,
Taiyuan, Shanxi 030006, China}

\thanks{\emph{e-mail:duyzh22@sxdtdx.edu.cn, kietemap@126.com, zhangyang1503575@163.com, gudujianghu23@163.com}}

\begin{abstract}
Due to quantum gravitational effects, Barrow proposed that the black hole horizon is ``fractalized'' into a sphereflake. Based on this issue, in this work we investigate the phase structure and stability of Einstein-Power-Yang-Mills (EPYM) AdS black holes in the restricted phase space, assuming the black hole event horizon has a fractal structure. From the first law of thermodynamics for EPYM AdS black hole in the restricted phase space, we find that the mass parameter should be interpreted as the internal energy. Moreover, the Smarr relation of this system in the restricted phase space is not a homogeneous function due to the fractal structure, which is fully differs significantly from the corresponding relation in the extended phase space. The presence of a fractal structure can be regarded as a probe for phase transitions. Interestingly, for a fixed central charge in EPYM AdS black hole system with a fractal structure, a supercritical phase transition also exists, similar to the case in the strandard EPYM AdS black hole system. Furthermore, the effects of the fractal parameter $\Delta$ and non-linear Yang-Mills parameter $\gamma$ on the thermodynamical stability of this system are also investigated.
\end{abstract}


%\pacs{04.70.Dy,05.70.Ce,04.20.Gz}
%\pacs{04.70.Dy 05.70.Ce}
\maketitle

\section{Introduction}
As is well known, black holes are both thermodynamic and quantum systems \cite{Bekenstein1973,Jacobson1995,Padmanabhan2010}. Black hole thermodynamics is a very important research field because it can provide some clues for exploring the nature of quantum gravity. Since Bekenstein and Hawking proposed the Bekenstein-Hawking entropy \cite{Bekenstein1973,Hawking1975} for black hole, the corresponding thermodynamic laws of black holes \cite{Bekenstein1973,Hawking1983} were presented by analogy with ordinary thermodynamic systems. Subsequently, the study of thermodynamic properties of black holes has garnered substantial interest. Among them, the most important is the investigation of the asymptotically anti-de Sitter (AdS) black hole thermodynamics. Though the AdS/CFT correspondence \cite{Maldacena1998} the thermodynamical properties of an AdS black hole with a finite temperature can be described by the dual conformal field theory (CFT). When an AdS black hole is in thermal equilibrium with its surrounded radiation field, the famous Hawking-Page (HP) phase transition \cite{Hawking1983,Witten1998} will survive. For the RN-AdS black hole, in $1999$ Chamblin found the first-order phase transition of this system \cite{Chamblin1999}. It should be noted that by identifying the negative cosmological constant to the positive thermodynamical pressure \cite{Kastor2009}, the traditional black hole thermodynamics was extended to the extended phase space where black hole mass is regarded as enthalpy. In the extended phase space AdS black holes have more and more richer phase structure and thermodynamical properties: the Van de Waals-like phase transition \cite{Kubiznak2012,Wei2015,Wei2024a}, the reentrant phase transitions \cite{Altamirano2013,Frassin2014}, superfluid \cite{Hennigar2017a}, the polymer-like phase transition \cite{Dolan2014}, the triple points \cite{Wei2014,Li2022}, the novel dual relation of HP phase transition \cite{Wei2020}, and the topological structure \cite{Wei2024,Wei2022a}. Furthermore, the propose of the extended phase space for AdS black holes reduce to the appearance of other novel dual thermodynamical quantities that related with model parameters \cite{Cai2013} and to the investigation of some black hole heat engines \cite{Johnsom2014,Xu2017}. All of those developments are in the subdiscipline, black hole chemistry \cite{Kubiznak2017}.

From the viewpoint of the holography \cite{Zhang2015,Dolan2016}, the explanation of the black hole chemistry via AdS/CFT \cite{Dolan2014a,Kastor2014} have somewhat elusive: i) For an AdS black hole, the changing of the cosmological constant $\delta\Lambda$ corresponds to both the changing of the central charge and the CFT volume. This reduces to the inconsistency of the thermodynamical first law between the extended phase space and the dual field theory \cite{Karch2015,Sinamuli2017}. ii) $\delta\Lambda$ (or $\delta P$) implies the changing of the gravity model. The corresponding ensemble does not describe the collection of black holes from the same gravity model where the micro states are the same, while describes the collection of gravity models of the same or similar black hole solutions. iii) The Smarr relation of an AdS black hole in the extended phase space is not a homogeneous function of order one for all independent dual thermodynamical quantities. That prompts people to seek for the more expanded first law of thermodynamics. Recently, the central charge and the chemical potential which are as a new pair dual thermodynamical quantities were introduced in the first law of thermodynamics \cite{Cong2021,Visser2022}, i.e., the extended phase space is extended to the restricted phase space. The corresponding Newton's gravitational constant and the cosmological constant both can change. And it can induce profound consequences of the chemical potential and its holographic interpretation. In the restricted phase space, there rise a new thermodynamical phenomenon of AdS black holes that is fully different from the first-order phase transition of AdS black hole in the extended phase space and ordinary thermodynamical systems, the supercritical phase transition \cite{Gao2021,Zhao2022,Sadeghi2022}. In this work, for the Einstein-Power-Yang-Mills (EPYM) AdS black hole \cite{Zhang2015a,Corda2011,Mazharimousavi2009,Lorenci2002} we will exhibit the concrete process of establishing the thermodynamical first law in the restricted phase space.

On the other hand, Black hole thermodynamics is the most important tool to investigate the quantum gravity in the spacetime containing the horizon \cite{Birrell1982}. The Hawking radiation of black hole provides a valuable insight that the Hawking entropy is proportional to the black hole horizon area and Hawking temperature is proportional to the black hole horizon surface gravity \cite{Abreu2020}. This indicates that there exists a strong relationship between thermodynamical systems and gravitational systems. In 2020 Barrow proposed that due to the quantum gravitational effects the surface of a black hole can be altered, which leads to the emergence of the Barrow entropy \cite{Barrow2020}. And the black hole horizon area can be represented as a discontinuous and possibly fractal structure of the geometry of the horizon. Based on the concept of the ``Koch snowflake'', Barrow thought the Schwarzschild black hole horizon is surrounded by $N$ smaller spheres whose radii have a ratio $¦Ë$ with respect to the original sphere. After the infinite number of steps the total area and volume are the sum of all the intricate structures, which are finite and infinite, respectively. Thence the corresponding entropy becomes very large. Subsequently, the idea of the Barrow entropy has been further used to the investigation of the dark energy \cite{Saridakis2020,Moradpour2020}, the cosmology \cite{Salehi2023,Komatsu2024,Okcu2024}, and black hole thermodynamics \cite{Abreu2020a,Wang2022,Ladghami2024,Rani2023}. In this work we will investigate the thermodynamics and stability of AdS black hole with the fractal structure of the black hole horizon.

Additionally, the linear charged black holes in AdS spacetime within a second-order phase transition show a scaling symmetry: at the critical point the state parameters scale respects to charge q, i.e., $S\sim q^2,~P\sim q^{-2},~T\sim q^{-1}$ \cite{Johnson2018}. It is naturally to gauss whether there exists the scaling symmetry in the non-linear charged AdS black holes. As a generalization of the charged AdS Einstein-Maxwell black holes, it is interesting to explore new non-linear charged systems. Due to infinite self-energy of point like charges in Maxwell's theory \cite{Born1934,Kats2007,Anninos2009,Cai2008,Seiberg1999}, Born and Infeld proposed a generalization when the field is strong, bringing in non-linearities \cite{Dirac2013,Birula1970}. An interesting non-linear generalization of charged black holes involves a Yang-Mill field exponentially coupled to Einstein gravity, i.e., the Einstein-power-Yang-Mills (EPYM) theory, which possesses the conformal invariance and is easy to construct the analogues of the four-dimensional Reissner-Nordstr\"{o}m black hole solutions in higher dimensions. Unlike the case of the Maxwell coupled to Einstein gravity whose range extends to infinity, considerations of the coupling between Einstein gravity and the Yang-Mills (YM) field inside the nuclei and highly dense matter systems are important justifying their inclusion in applications to black holes. Corresponding several features of the EPYM gravity in extended thermodynamics have recently been studied \cite{Du2021,Zhang2015,Moumni2018}. On the holographically dual side, effects of the nonlinear sources on the strongly coupled dual gauge theory in the context of AdS/CFT correspondence have also been reported \cite{Roychowdhury2013}.

This work is organized as follows. In Sec. \ref{scheme2}, we give a brief description of the EPYM AdS black hole in the restricted phase space. In Sec. \ref{scheme3}, we exhibit the phase structure of the EPYM AdS black hole with the fractal structure on the black hole horizon in the restricted phase space. Then, the stability of the EPYM AdS black hole with the fractal structure on the black hole horizon in the restricted phase space is analyzed in Sec. \ref{scheme4}. Furthermore, the effects of the fractal parameter and the non-linear YM charge parameter on the phase structure and stability are also probed. A brief summary is given in Sec. \ref{scheme5}.

\section{A brief description of Barrow EPYM AdS black hole in the restricted phase space}
\label{scheme2}
For the four-dimensional AdS Einstein-power-Yang-Mills (EPYM) black hole, the spherically symmetric static spacetime's parameterization approach is as follows
\begin{eqnarray}
d s^{2}=-f(r) d t^{2}+f^{-1} d r^{2}+r^{2} d \Omega_{2}^{2},
\end{eqnarray}
the metric function can be described as \cite{Yerra2018}
\begin{eqnarray}
f(r)=1-\frac{2GM}{r}+\frac{r^{2}}{l^2}+\frac{G\left(2q^{2}\right)^{\gamma}}{2(4 \gamma-3) r^{4 \gamma-2}}. \label{f}
\end{eqnarray}
Here $d\Omega_{2}^{2}$ is the metric on unit $2$-sphere with volume $4\pi$ and $q$ is the YM charge, $l$ is related to the cosmological constant: $l^2=-\frac{3}{\Lambda}$, $G$ is the gravitational constant, $\gamma$ is the non-linear YM charge parameter and satisfies $\gamma>0$ \cite{Corda2011}. Additionally, the event horizon $r_+$ corresponds to the largest root by solving the expression $f(r_+)=0$. Thus the mass parameter of the black hole can be expressed in terms of the horizon radius as
\begin{eqnarray}
M=\frac{r_+}{2G}\left(1+\frac{r^2_+}{l^2}
+\frac{2^{\gamma-1}Gq^{2\gamma}}{(4\gamma-3)r_+^{4\gamma-2}}\right).
\label{M}
\end{eqnarray}
From the classical Boltzmann-Gibbs statistics the famous Bekenstein-Hawking entropy is proportional to the surface area of the black hole horizon: $S=\pi r_+^2/G$. Here we should stress that although $S$ commonly used in the relativistic theory, it is still essentially classical. From the maximum entropy principle, it is maximized when a thermodynamic equilibrium system is described by the Maxwell-Boltzmann distribution which is in the classical Boltzmann-Gibbs statistics framework. Note that arguments from multiple perspectives indicate the Boltzmann-Gibbs statistics maybe not the appropriate theme for investigating the black hole thermodynamics.

In Ref. \cite{Barrow2020} Barrow proposed that the quantum-gravitational effect maybe arises a complex structure of black hole horizon surface extending to microscopic scales. These effects manifest as changes in the black hole horizon area that a fractal structure can describe. By considering the impact of quantum gravity and the fractal geometry Barrow alters the conventional perception of a smooth and uniform event horizon. This fractal characteristic implies that the event horizon undergoes complex transformations. Indeed, thermodynamics states that black holes are infinitely complex systems, but by these changes in the black hole horizon area as a limited volume but an infinite/finite surface area, a new entropy relation will arise,
called Barrow entropy, which reads as follows
\begin{eqnarray}
S_B=(\pi r_+^2/G)^{1+\Delta/2}, \label{SB}
\end{eqnarray}
where the parameter $\Delta$ stands for the quantum effect and it yields  $(\Delta\in[0,1])$. As $\Delta=0$ the well-known Bekenstein-Hawking entropy is restored, signifying the absence of a fractal structure. Conversely, when $\Delta=1$ it indicates the most deformed and complex fractal structure of the black hole horizon surface. Besides, the Barrow entropy is associated with Tasllis entropy execution \cite{Tsallis1988,Tsallis2013}, and it is unique from the quantum corrected entropy with logarithmic \cite{Saridakis2020,Nozari2007,Nozari2006}. Furthermore, there exist other entropy functions such as the Renyi entropy \cite{Neyman1960}, the Sharma-Mittal entropy \cite{Jahromi2018}, and the Kaniadakis entropy \cite{Kaniadakis2005}. In the following we will investigate the Barrow entropy of the EPYM AdS black hole and its thermodynamics in the restricted phase space. We can also obtain the Barrow temperature of the EPYM AdS black hole from eqs. (\ref{M}) and (\ref{SB}) as follows
\begin{eqnarray}
\frac{1}{T_B}=\frac{\partial S_B}{\partial M}=\frac{2\pi(2+\Delta)(\pi/G)^{\Delta/2} r_{+}^{1+\Delta}}{1+3r_{+}^{2}/l^2-\frac{G\left(2 q^{2}\right)^{\gamma}}{2 r_{+}^{4 \gamma-2}}}.\label{T}
\end{eqnarray}
In the extended phase space (setting $G=1$), the thermodynamical first law for the EPYM AdS black hole reads
\begin{eqnarray}
d M&=&Td S+Vd P+\bar\Phi d q^{2\gamma},\label{deltaM}
\end{eqnarray}
considering eq. (\ref{M}) and $P=\frac{3}{8\pi l^2}$ , the corresponding thermodynamical volume and potential are
\begin{eqnarray}
V=\frac{\partial M}{\partial P}=\frac{4\pi r_+^3}{3},~~~~
\bar\Phi=\frac{\partial M}{\partial q^{2\gamma}}=
\frac{2^{\gamma-2}}{(4\gamma-3)r_+^{4\gamma-3}}.\label{VPsi}
\end{eqnarray}
The corresponding thermodynamical properties of the EPYM AdS black hole in the extended phase space were exhibited in Refs. \cite{Du2021,Du2023} and the optical properties such as the photon sphere and shadow were also presented in Refs. \cite{Du2022,Du2022a}.

Note that from the Holographic interpretation of the thermodynamical first law (\ref{deltaM}) in the extend phase space there exist some issues \cite{Johnsom2014,ZhangJHEP2015,McCarthyJHEP2017}. From the boundary conformal field theory (CFT), the term of $V\delta P$ can arise to two terms in the thermodynamical first law: one is the central charge of the boundary CFT, the other one is the thermodynamical pressure of the boundary CFT which is caused by the change of the AdS radius. The way of addressing this problem is to invoke the form of the central charge from the AdS/CFT dictionary, which is related to the AdS radius $l$ as in Ref. \cite{Karch2015}
\begin{eqnarray}
C=\frac{l^2}{G}.\label{C}
\end{eqnarray}
The partition function of AdS spacetime is related to the Euclidean action \cite{Gibbons1977,Chamblin1999a} by

\begin{eqnarray}
I_E=-\ln Z_{AdS},
\end{eqnarray}
where for the EPYM AdS black hole $I_E=\frac{1}{2}\int d^4x\sqrt{g}\left(R-2\Lambda-\mathcal{F}^\gamma\right)$. The Yang-Mills (YM) invariant $\mathcal{F}$ and the YM field $F_{\mu \nu}^{(a)}$ read
\begin{eqnarray}
\mathcal{F}&=&\operatorname{Tr}(F^{(a)}_{{\mu\nu}}F^{{(a)\mu\nu}}),~~~~\operatorname{Tr}(F^{(a)}_{\mu\nu}F^{(a)\mu\nu})
=\sum_{a=1}^3 F^{(a)}_{\mu\nu}F^{(a)\mu\nu},\\
F_{\mu \nu}^{(a)}&=&\partial_{\mu} A_{\nu}^{(a)}-\partial_{\nu} A_{\mu}^{(a)}+\frac{1}{2 \xi} C_{(b)(c)}^{(a)} A_{\mu}^{(b)} A_{\nu}^{(c)}.
\end{eqnarray}
Here $R$ and $\gamma$ are the scalar curvature and a positive real parameter, respectively; $C_{(b)(c)}^{(a)}$ represents the structure constants of three-parameter Lie group $G$; and $\xi$ is the coupling constant. $A_{\mu}^{(a)}$ are the $SO(3)$ gauge group Yang-Mills (YM) potentials defining by the Wu-Yang (WY) ansatz \cite{Balakin2016}. Note that the internal indices ${a,b,c,...}$ do not differ whether in covariant or contra-variant form. Implementing the variation of the action with respect to the spacetime metric $g_{\mu\nu}$, the gravitational the field equations yields
\begin{eqnarray}
&&G^{\mu}{ }_{\nu}+\Lambda\delta^{\mu}{ }_{\nu}=T^{\mu}{ }_{\nu},\\
&&T^{\mu}{ }_{\nu}=-\frac{1}{2}\left(\delta^{\mu}{ }_{\nu} \mathcal{F}^{\gamma}-4 \gamma \operatorname{Tr}\left(F_{\nu \lambda}^{(a)} F^{(a) \mu \lambda}\right) \mathcal{F}^{\gamma-1}\right).
\end{eqnarray}
Through the variation with respect to the YM gauge potentials $A_{\mu}^{(a)}$ and the traceless condition, the $2$-forms YM equations yields
\begin{equation}
\mathbf{d}\left({ }^{\star} \mathbf{F}^{(a)} \mathcal{F}^{\gamma-1}\right)+\frac{1}{\xi} C_{(b)(c)}^{(a)} \mathcal{F}^{\gamma-1} \mathbf{A}^{(b)} \wedge^{\star} \mathbf{F}^{(c)}=0,
\end{equation}
where $\mathbf{F}^{(a)}=\frac{1}{2}F_{\mu \nu}^{(a)}dx^\mu\wedge dx^\nu,~\mathbf{A}^{(b)}=A_{\mu }^{(b)}\wedge dx^\mu$, and ${ }^{\star}$ stands for duality. It is obviously that for the case of $\gamma=1$ the EPYM theory reduces to the standard Einstein-Yang-Mills (EYM) theory \cite{Mazharimousavi2007}. Note that the non-Abelian property of the YM gauge field is expressed with its YM potentials
\begin{eqnarray}
\mathbf{A}^{(b)}=\frac{q}{r^2}C^{(a)}_{(i)(j)}x^idx^j,~r^2=\sum_{j=1}^3x_j^2,
\end{eqnarray}
and $q$ is the YM charge, the indices ($a,~i,~j$) obey the following ranges: $1\leq a,~i,~j\leq3$. The coordinates $x_i$ take the following forms: $x_1=r \cos\phi \sin\theta,~x_2=r \sin\phi \sin\theta,~x_3=r \cos\theta.$ Since we have utilized the WY ansatz for the YM field, the invariant for this field takes the form \cite{Stetsko2020,Chakhchi2022}
\begin{eqnarray}
\operatorname{Tr}(F^{(a)}_{{\mu\nu}}F^{{(a)\mu\nu}})=\frac{q^2}{r^4}.
\end{eqnarray}
This form leads to the disappearance of the structure constants which can be described the non-Abelian property of the YM gauge field. Therefore, under the condition of the WY ansatz we may focus on the role of the non-linear YM charge parameter, instead of the non-Abelian character parameter. Furthermore, in order to guarantee the Weak Energy Condition (WEC), Strong Energy Condition (SEC), Dominant Energy Condition (DEC), and Causality Condition (CC) of this system, the physically meaningful range of the non-linear YM charge parameter $\gamma$ should be bigger than $3/4$ and less than $3/2$ (see Ref. \cite{Mazharimousavi2009}).

From the formulation of the AdS/CFT correspondence, the partition function of AdS spacetime is equal the CFT dual one, $Z_{AdS}=Z_{CFT}$. The partition function of CFT is related to the free energy by
\begin{eqnarray}
\bar{G}=-T \ln Z_{CFT},
\end{eqnarray}
where $T$ stands for the temperature. As shown in Ref. \cite{Visser2022}, the free energy is in terms of the central charge and the chemical potential as follows
\begin{eqnarray}
\bar{G}=\mu C.
\end{eqnarray}
One can infer that the free energy is correlated with the parameters characterizing EPYM AdS black hole
\begin{eqnarray}
\mu C=T I_E=M-TS-\Phi Q.
\end{eqnarray}
Here $S$ corresponds to the Bekenstein-Hawking entropy, $\Phi$ is the electric potential on the black hole event horizon due to the electric charge $Q=q^{\gamma}\sqrt{C}$. The corresponding Smarr relation within the restricted phase space can be written as
\begin{eqnarray}
M=TS+\Phi Q+\mu C. \label{MC}
\end{eqnarray}
The first law of thermodynamic for the EPYM AdS black hole in the restricted phase space reads as follows (more details see Ref. \cite{Du2023a})
\begin{eqnarray}
dM=TdS+\Phi dQ+\mu dC.
\end{eqnarray}
Since $S_B=S^{1+\Delta/2}$ and the Hawking temperature $T=\partial M/\partial S$, one can establish the relationship between the modified quantities ($T_B,~S_B$) and the ordinary ones ($T,~S$) as
\begin{eqnarray}
TS=(1+\Delta/2)T_BS_B.
\end{eqnarray}
Substituting above equation into the eq. (\ref{MC}), we can obtain the Smarr relation in terms of the modified quantities
\begin{eqnarray}
M=(1+\Delta/2)T_BS_B+ \Phi  Q+\mu C \label{MCC}.
\end{eqnarray}
This results contradicts the nature of the Smarr relation in the restricted phase space in the normal case, where the mass parameter is a homogeneous function of order one for all quantities. The effects of quantum gravity represented by the fractal parameter $\Delta$ destroy the homogeneous of Smarr relation in the restricted phase space. As $\Delta=0$ the eq. (\ref{MCC}) can recover the Smarr relation of the ordinary restricted phase space. With the definition of Barrow temperature $T_B=\frac{\partial M}{\partial S_B}$, one can obtain $T_B=\frac{\partial M}{\partial S}\frac{\partial S}{\partial S_B}=T\frac{\partial S}{\partial S_B}$, i.e., $TdS=T_BdS_B$. Thus the corresponding thermodynamics first law for the Barrow EPYM AdS black hole in the restricted phase space becomes as
\begin{eqnarray}
dM=T_B dS_B+ \Phi d Q+\mu dC.
\end{eqnarray}

\section{Phase structure of Barrow EPYM AdS black hole in the restricted phase space}
\label{scheme3}

\begin{figure}[htp]
\centering
\subfigure[$~T_{Bc}-C_c$]{\includegraphics[width=0.4\textwidth]{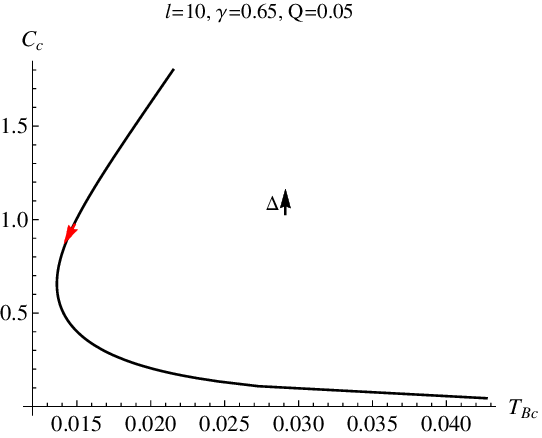}\label{Cc-TBc-delta}}~~~~
\subfigure[$~T_{Bc}-C_c$]{\includegraphics[width=0.4\textwidth]{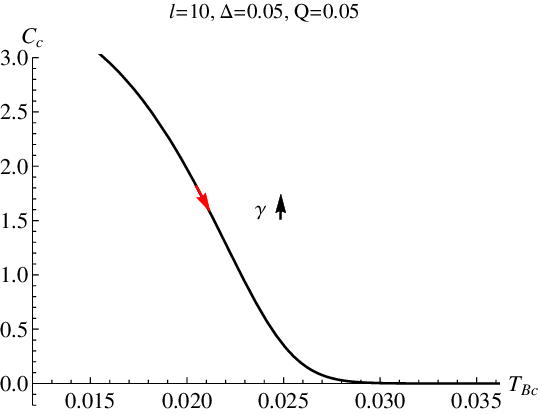}\label{Cc-TBc-gamma}}
\caption{The behaviors of the center charge respects to the Barrow's temperature $T_{Bc}$ with different non-linear charge parameter $\gamma$ and the fractal parameter $\Delta$. }\label{Cc-TBc-delta-gamma}
\end{figure}

In this part, we will investigate the thermodynamical properties of the Barrow EPYM AdS black hole in the restricted phase space. First, we establish the thermodynamical quantities and the equations of state. With eqs. (\ref{M}), (\ref{SB}), and (\ref{C}), the black hole mass parameter in terms of the modified quantities and the fractal parameter becomes
\begin{eqnarray}
M(S_B,C, Q,\Delta)=\frac{\sqrt{C}S_B^{\frac{1}{2+\Delta}}}{2l\sqrt{\pi}}
\left(1+\frac{S_B^{\frac{2}{2+\Delta}}}{\pi C}
+\frac{(2\pi^2C^2)^\gamma l^{4-4\gamma} Q^2}{2(4\gamma-3)\pi C^2 S_B^{\frac{4\gamma-2}{2+\Delta}}}\right).
\end{eqnarray}
Since $T_B=\left(\frac{\partial M}{\partial S_B}\right)_{C, Q,\Delta}$ and $\mu=\left(\frac{\partial M}{\partial C}\right)_{S_B, Q,\Delta}$, the equations of state describing the EPYM AdS black hole in the presence of the quantum gravity effects are as follows
\begin{eqnarray}
T_B&=&\frac{\pi CS_B^{\frac{4\gamma-2}{2+\Delta}}+3S_B^{\frac{4\gamma}{2+\Delta}}-2^{\gamma-1}\pi(\pi C)^{2\gamma-2}l^{4-4\gamma} Q^2}{2(2+\Delta)\sqrt{C}\pi^{3/2}lS_B^{\frac{4\gamma-1+\Delta}{2+\Delta}}},\label{TBSB}\\
\mu&=&\frac{S_B^{\frac{1}{2+\Delta}}}{4l\sqrt{\pi C}}
\left(1-\frac{S_B^{\frac{2}{2+\Delta}}}{\pi C}
+\frac{(4\gamma-1)(2\pi^2C^2)^\gamma l^{4-4\gamma}\bar Q^2}{2(4\gamma-3)\pi C S_B^{\frac{4\gamma-2}{2+\Delta}}}\right).
\end{eqnarray}
Based on the classification of phase transition for a thermodynamical system by Ehrenfest, the critical point can be obtained by the following equations
\begin{eqnarray}
\frac{\partial T_B}{\partial S_B}=\frac{\partial^2T_B}{\partial S_B^2}=0.
\end{eqnarray}
Considering the eq. (\ref{TBSB}) we can obtain the critical thermodynamical quantities as follows
\begin{eqnarray}
%r_{c}^{4-4\gamma}&=&\frac{6C_c(1-\Delta)}{(2\gamma-1)(4\gamma-1+\Delta)2^{\gamma}q^{2\gamma}},~~
C_{c}&=&\frac{\gamma(4\gamma-1+\Delta)2^{\gamma}Q^{2}l^{4-4\gamma}}{1+\Delta}
\left(\frac{6\gamma(1-\Delta)}{(2\gamma-1)(1+\Delta)}\right)^{2\gamma-1},\label{Cc}\\
S_{Bc}^{\frac{2}{2+\Delta}}&=&\frac{\pi C_c(1+\Delta)(2\gamma-1)}{6\gamma(1-\Delta)},~~
T_{Bc}=\frac{\pi C_cS_{Bc}^{\frac{4\gamma-2}{2+\Delta}}+3S_{Bc}^{\frac{4\gamma}{2+\Delta}}-2^{\gamma-1}\pi(\pi C_c)^{2\gamma-1}l^{4-4\gamma} Q^2}{2(2+\Delta)\sqrt{C_c}\pi^{3/2}lS_{Bc}^{\frac{4\gamma-1+\Delta}{2+\Delta}}}.
\end{eqnarray}
The results indicate that the critical point is determined by the non-linear YM parameter $\gamma$, the fractal parameter $\Delta$, the AdS spacetime radius $l$, and the electric charge $Q$. The corresponding behaviour of the critical center charge with respect to the critical Barrow's temperature under different parameters have been depicted in Fig. \ref{Cc-TBc-delta-gamma}. When the parameters ($l,~\gamma,~Q$) are fixed, the critical center charge increases with the critical Barrow's temperature for the small value of the fractal parameter $\Delta$, and for the large values of $\Delta$ it decreases with $T_{Bc}$, see Fig. \ref{Cc-TBc-delta}. For small values of the non-linear YM charge parameter $\gamma$ the critical center charge monotonically and rapidly decays with the critical Barrow's temperature until to zero, while it tends to zero and does not changes with the critical Barrow's temperature, see Fig. \ref{Cc-TBc-gamma}.
%%%%%%%%%%%%%%%%%%%%%%%%%%%%%%%%%%%%%%%%%%%%%%%%%%%%%%%%%%%%%%%%%%%%%%%%%%%%%%%%%%%%%%%%%%%%%%%%%%

\begin{figure}[htp]
\centering
\includegraphics[width=0.4\textwidth]{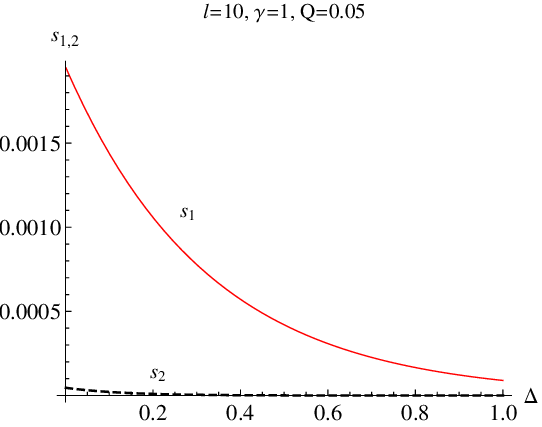}
\caption{The behaviors of the center charge respects to the Barrow's temperature $T_{Bc}$ with different non-linear charge parameter $\gamma$ and the fractal parameter $\Delta$. }\label{s12-delta}
\end{figure}

In order to investigate the first-order phase transition and obtain the analyzed results, we consider the case of $\gamma=1$ to solve the equation $\frac{\partial T_B}{\partial S_B}=0$. With the definitions $s=S_B/S_{Bc},~c=C/C_c$, there are two solutions of the normalized Barrow entropy which read as follows
\begin{eqnarray}
s_{1,2}^{\frac{2}{2+\Delta}}=
c\left(1\pm\sqrt{1-\frac{24Q^2(1-\Delta)(3+\Delta)}{(1+\Delta)^2}}\right),
\end{eqnarray}
where $s_1$ stands for the solution of $`+'$ and $s_2$ is the solution of $`-'$. Note that these two solutions are related with the first-order phase transition and the behaviours of them respect to the fractal parameter are depicted in Fig. \ref{s12-delta}. We can see that with the increasing of the fractal parameter $\Delta$ both two dimensionless solutions also decrease. When the fractal structure reaches to the maximum $\Delta=1$, $s_1=\sqrt{2c}$ and $s_2=0$. Here we should point out that for the maximum fractal structure there exists only one case of the critical point, $s_c=s_{1,2}=0$, which is meaningless and can be ruled out. Additionally, when this system is without the fractal structure $\Delta=0$, $s_1=s_2=c=1$ as $Q^2=1/36$. This case is consistent with that of the EPYM AdS black in the restricted phase space with $\gamma=1$ (see Ref. \cite{Du2023a}). Note that at the critical point we have $Q^2=q^2/C_c=1/36$ as $\gamma=1$, i.e., $q^2=C_c/36$. That is just the relation between the YM charge and the critical center charge in eq. (\ref{Cc}). From the above analysis, it can be seen the fractal structure caused by the quantum gravity plays an important role in the critical phenomena and first-order phase transition: when the fractal structure reaches to the maximum, there is no phase transition can been survived. Hence, due to the quantum gravity effects on the black hole horizon surface the fractal structure behaves as a phase transition probe.
%%%%%%%%%%%%%%%%%%%%%%%%%%%%%%%%%%%%%%%%%%%%%%%%%%%%%%%%%%%%%%%%%%%%%%%%%%%%%%%%%%%%%%%%%%%%%%%%%%

\begin{figure}[htp]
\centering
\subfigure[$~s-t$]{\includegraphics[width=0.4\textwidth]{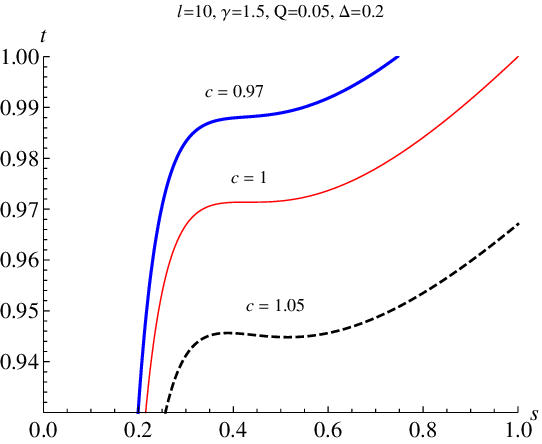}}~~
\subfigure[$~s-t$]{\includegraphics[width=0.4\textwidth]{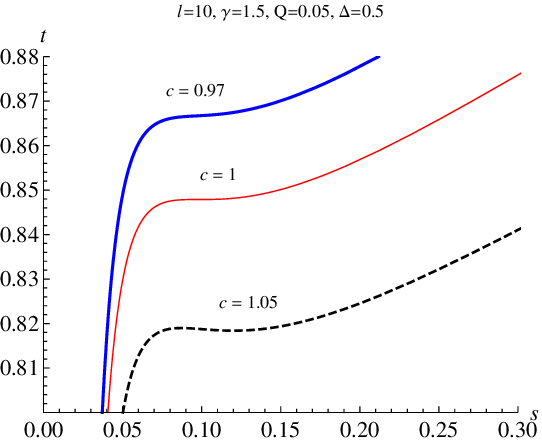}}~~
\caption{The phase diagram of $t-s$ with different normalized center charge $c$ and the fractal parameter $\Delta$. }\label{s-t-c}
\end{figure}

\begin{figure}[htp]
\centering
\subfigure[$~t-g$]{\includegraphics[width=0.4\textwidth]{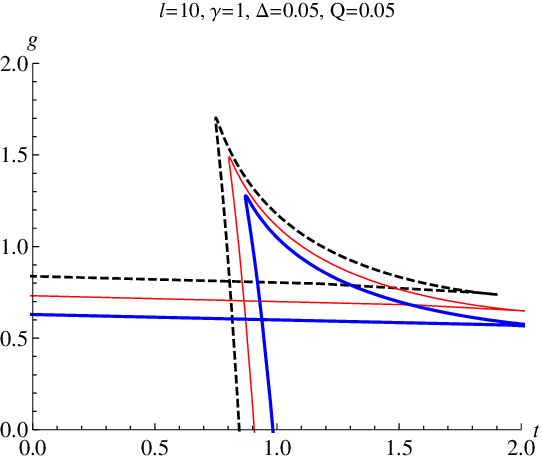}}~~
\subfigure[$~t-g$]{\includegraphics[width=0.4\textwidth]{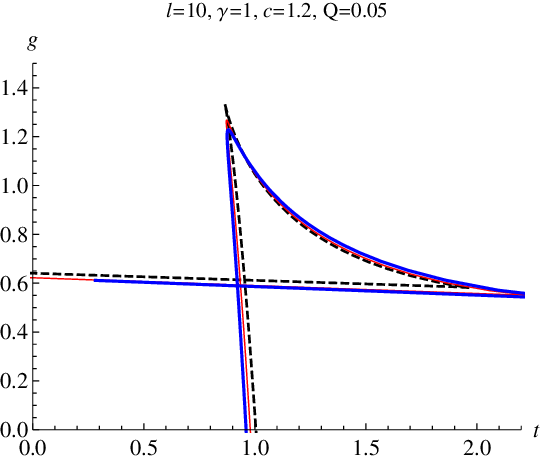}}
\caption{The first-order phase transition in the phase diagram of $t-g$ with different normalized center charge $c$ and the fractal parameter $\Delta$. In the left file, the parameter is set to $c=1.2$ (the blue thick line), $c=1.4$ (the red thin line), and $c=1.6$ (the dashed black line); in the right file, the fractal parameter is set to $0$ (the blue thick line), $0.06$ (the red thin line), and $0.11$ (the dashed black line).}\label{t-g-c}
\end{figure}

On the one hand we can investigate the first-order phase transition from the perspective of the Maxwell's equal area law in $T_B-S_B$ phase diagram: $\int_{S_{B1}}^{S_{B2}} T_B dS_B=T_{B0}(S_{B2}-S_{B1})$ with the first-order phase temperature $T_{B0}$ and the boundary entropies $ S_{B1,B2}$ in the two-phases coexistence. Since there is no analytic expression for different values of $\Delta$ and $\gamma$, here we will only numerically give the corresponding phase diagram. For convenience, with the normalized parameter $t=T_B/T_{Bc}$, one can obtain
\begin{eqnarray}
t=\frac{1+3(sS_{Bc})^{\frac{2}{2+\Delta}}-2^{\gamma-1}Q^2l^{4-4\gamma}\pi^{2\gamma-1}
(cC_c)^{2\gamma-2}(sS_{Bc})^{-\frac{4\gamma-2}{2+\Delta}}}
{s^{\frac{1+\Delta}{2+\Delta}}
\left(1+3S_{Bc}^{\frac{2}{2+\Delta}}-2^{\gamma-1}Q^2l^{4-4\gamma}\pi^{2\gamma-1}
C_c^{2\gamma-2}S_{Bc}^{-\frac{4\gamma-2}{2+\Delta}}\right)}.
\end{eqnarray}
From the above equation we plot the normalized Barrow's temperature as a function of the normalized Barrow entropy with different parameters in Fig. \ref{s-t-c}.
%%%%%%%%%%%%%%%%%%%%%%%%%%%%%%%%%%%%%%%%%%%%%%%%%%%%%%%%%%%%%%%%%%%%%%%%%%%%%%%%%%%%%%%%%%%%%%%%%%
The results show that for the Barrow EPYM AdS black hole in the restricted phase space, there exists the supercritical phenomenon of the phase transition: when $c<1$, it is a monotonic curve of $s-t$ in which no phase transition occurs; as $c=1$ the critical point appears; this system undergoes a first-order phase transition as $c>1$. On the other hand, from the Gibbs free energy
\begin{eqnarray}
g=\bar{G}/\bar{G}_c=\sqrt{c}s^{\frac{1}{2+\Delta}}
\frac{1-\frac{1}{\pi cC_c}(sS_{Bc})^{\frac{2}{2+\Delta}}
+\frac{(4\gamma-1)2^{\gamma-1}Q^2l^{4-4\gamma}\pi^{2\gamma-1}
(cC_c)^{2\gamma-1}}{(4\gamma-3)(sS_{Bc})^{\frac{4\gamma-2}{2+\Delta}}}}
{1-\frac{1}{\pi C_c}S_{Bc}^{\frac{2}{2+\Delta}}
+\frac{(4\gamma-1)2^{\gamma-1}Q^2l^{4-4\gamma}\pi^{2\gamma-1}
C_c^{2\gamma-1}}{(4\gamma-3)S_{Bc}^{\frac{4\gamma-2}{2+\Delta}}}},
%f=F/F_c=\sqrt{c}s^{\frac{1}{2+\Delta}}
%\frac{1-\frac{1-\Delta}{\pi cC_c(1+\Delta)}(sS_{Bc})^{\frac{2}{2+\Delta}}
%+2^{\gamma-1}Q^2l^{4-4\gamma}\pi^{2\gamma-1}
%(cC_c)^{2\gamma-2}(sS_{Bc})^{-\frac{4\gamma-2}{2+\Delta}}}
%{1-\frac{1-\Delta}{\pi C_c(1+\Delta)}S_{Bc}^{\frac{2}{2+\Delta}}
%+2^{\gamma-1}Q^2l^{4-4\gamma}\pi^{2\gamma-1}
%C_c^{2\gamma-2}S_{Bc}^{-\frac{4\gamma-2}{2+\Delta}}}, ~~
\end{eqnarray}
the corresponding phase structure is also exhibited in Fig.\ref{t-g-c}. At the first-order phase transition point, the temperature of the Barrow EPYM AdS black hole in the restricted phase space decreases with the increasing of the normalized center charge $c$, while for smaller values of the fractal parameter $\Delta$ the phase transition temperature increases with $\Delta$. Therefore, it can be concluded that within the restricted phase space with the effect of quantum gravity on the event horizon area, the thermodynamics of the EPYM AdS black hole exhibit similarities to that of Van der Waals systems. Namely, in the phase diagram of temperature$-$entropy the monotonic curves mean that there is no phase transition, while the non-monotonic and multi-valued curves indicate the first-order phase transitions. From the viewpoint of Gibbs free energy, the swallow tail behaviors represent the existence of first-order phase transitions and the vertices represent the critical points.
%%%%%%%%%%%%%%%%%%%%%%%%%%%%%%%%%%%%%%%%%%%%%%%%%%%%%%%%%%%%%%%%%%%%%%%%%%%%%%%%%%%%%%%%%%%%%%%%%%

\section{Stability of Barrow EPYM AdS black hole in restricted phase space}
\label{scheme4}
\begin{figure}[htp]
\centering
\subfigure[$~s-\eta$]{\includegraphics[width=0.4\textwidth]{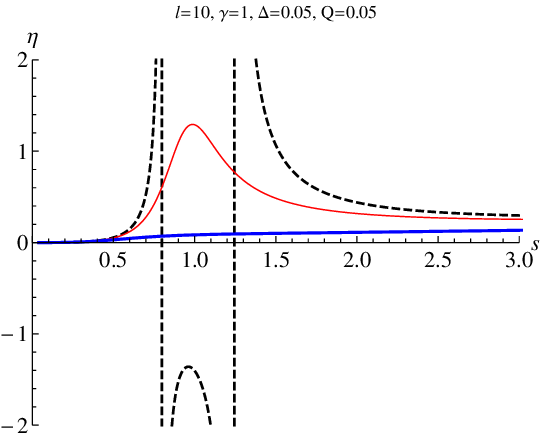}}\label{s-eta-c1}~~
\subfigure[$~s-\eta$]{\includegraphics[width=0.4\textwidth]{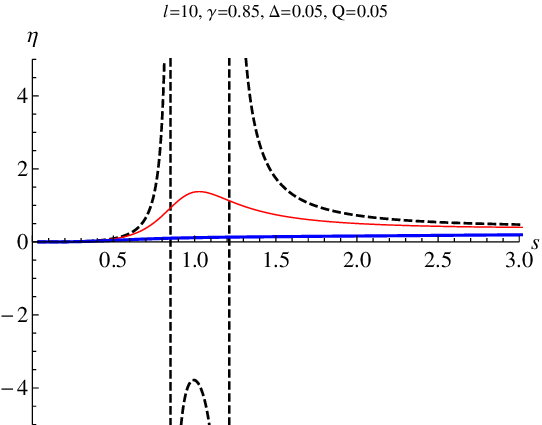}}\label{s-eta-c085}\\
\subfigure[$~s-\eta$]{\includegraphics[width=0.4\textwidth]{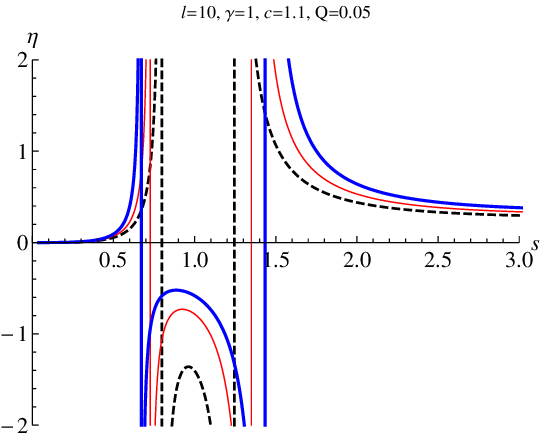}}\label{s-eta-delta}~~
\caption{The heat capacity of Barrow EPYM AdS black hole in the restricted phase space. In above two figures the normalized center charge is set to $c=1.1$ (the dashed black lines), $c=1$ (the red thin lines), and $c=0.5$ (the blue thick lines). In the down figure the fractal parameter is set to $\Delta=0.05$ (the dashed black line), $\Delta=0.1$ (the red thin line), and $\Delta=0.15$ (the blue thick line). }\label{s-eta}
\end{figure}

In the black hole thermodynamics, one key issue is to investigate the stability of black holes, which can be probed by the heat capacity. The positive heat capacity indicates a stable black hole, while the negative one stands for a instability black hole. In order to investigate the stability of Barrow EPYM AdS black hole in the restricted phase space and the effects of the fractal parameter $\Delta$ and the non-linear YM charge parameter $\gamma$ on the stability, one can obtain the corresponding heat capacity with the constant parameters ($l,~Q$) as the following form
\begin{eqnarray}
\eta=T_B\frac{\partial S_B}{\partial T_B}=-\frac{s(\Delta+1)S_{Bc}}{\Delta+2}\frac{1-\frac{\pi  2^{\gamma-1} Q^2 l^{4-4 \gamma} (\pi  c C_c)^{2 \gamma-2}}{(s S_{Bc})^{\frac{4\gamma-2}{\Delta+2}}}+\frac{3 (s S_{Bc})^{\frac{2}{\Delta+2}}}{\pi  c C_c}}{ 1-\frac{(3 (1-\Delta)) (s S_{Bc})^{\frac{2}{\Delta+2}}}{\pi  c C_c(\Delta+1)}-\frac{\pi  2^{\gamma-1} Q^2 (\pi C_c)^{2\gamma-2} (\Delta+4\gamma-1) l^{4-4 \gamma}}{(s S_{Bc})^{\frac{4\gamma-2}{\Delta+2}}}},
\end{eqnarray}
whose behaviors with different parameters $\Delta$ and $\gamma$ are shown in Fig. \ref{s-eta}.
We can conclude that when $c>1$, the EPYM AdS black hole in the restricted phase space with the fractal structure of the black hole horizon is undergoing a first-order phase transition among the large/intermediate/small black hole phases. The corresponding heat capacities of the large/small black hole phases are positive and they are stable, while the heat capacity of the intermediate one is negative and it is unstable. As $c=1$, i.e., a second-order phase transition between the large/small black hole phases, the heat capacities are positive and the corresponding black hole systems are stable. When $c<1$ this black hole system is of the positive heat capacity and one single stable phase. Furthermore, with the increasing of the non-linear YM charge parameter $\gamma$ and the fractal parameter $\Delta$, the region of the unstable intermediate phase also expands, the value of heat capacity also increases as $c\le1$.

\section{Discussions and Conclusions}
\label{scheme5}
In this manuscript we have studied the thermodynamics of the EPYM AdS black hole in the restricted phase space, considering the fractal structure of the black hole horizon. We have uncovered the influence of the quantum gravity and its impact on the black hole thermodynamics. The results reveal several remarkable characters, similar to those of the RN-AdS black hole in the restricted phase transition \cite{Gao2021}. Compared to the RN-AdS black hole in the expanded phase space, the results are summarized as followings:
\begin{itemize}
\item{The mass parameter is to be understood as the internal energy. Due to the fractal structure of the black hole horizon, the Smarr relation of this black hole in the restricted phase space is not restored, unlike in an ordinary thermodynamical system;}
\item{The EPYM AdS black hole with the fractal structure in the restricted phase space exhibits a unique thermodynamic phenomenon: the supercritical phase transition. When the system has a higher center charge, a first-order phase transition appears; however, it vanishes when the normalized center charge is less than one. This supercritical phase transition is governed by the degree of freedom in the conformal field theory;}
\item{When the fractal structure is maximum ($\Delta=1$), regardless of the normalized central charge parameter values, there is no physical phase transition in the EPYM AdS black hole. For the other values of the fractal structure parameter $0\leq\Delta<1$, there are non-zero thermodynamical quantities at the critical and first-order phase points. Hence the fractal structure can be regarded as a phase transition probe;}
\item{When $\Delta=0$, the phase transition property of the EPYM AdS black hole in the restricted phase space are the same as those in the extended phase space. This means that when a black hole lacks the fractal structure, its thermodynamical properties are independent of the choice of the corresponding phase spaces;}
\item{From the perspective of heat capacity, the EPYM AdS black hole in the restricted phase space with a black hole horizon with a fractal structure is always stable when the normalized center charge is equal to or less than one. In contrast, for $c>1$, this system is of one unstable intermediate black hole phase and two stable large/small black hole phases. These results are consistent with those in the extended phase space, where there is no fractal structure on the black hole horizon. Specifically, when the EPYM AdS black hole systems undergoes a first-order phase transition, the heat capacities of the two stable large/small black hole phases are always positive, while the heat capacity of the unstable intermediate black hole phase is negative. However, when the black hole systems is in no phase transition, the corresponding heat capacity is always positive (more details see Ref. \cite{Du2024}). }
\end{itemize}

\section*{Acknowledgements}
We would like to thank Prof. Ren Zhao and Meng-Sen Ma for their indispensable discussions and comments. This work was supported by the National Natural Science Foundation of China (No. 12075143), the Natural Science Foundation of Shanxi Province (No. 202303021211180).

\end{document}